\documentclass[letterpaper, prl, notitlepage,twocolumn]{revtex4-1}
\usepackage{latexsym}
\usepackage{graphicx}
\usepackage{color}
\begin{document}
\author{L.~W.~Kastens, S.~B.~Cahn, A.~Manzur, and D.~N.~McKinsey}
\affiliation{Department of Physics, Yale University, P.O. Box 208120, 
New Haven, CT 06520}
\title{Calibration of a Liquid Xenon Detector with $^{83}$Kr$^m$}

\begin{abstract}
We report the preparation of a $^{83}$Kr$^m$ source and its use in calibrating a liquid xenon detector. $^{83}$Kr$^m$ atoms were produced through the decay of $\rm ^{83}Rb$ and introduced into liquid xenon. Decaying $^{83}$Kr$^m$ nuclei were detected through liquid xenon scintillation. 
Conversion electrons with energies of 9.4 keV and 32.1 keV from the 
decay of $^{83}$Kr$^m$ were both observed. This calibration source will allow the
characterization of the response of
noble liquid detectors at low energies. $^{83}$Kr$^m$ may also
be useful for measuring fluid flow dynamics, both to understand
purification in noble liquid-based particle detectors, as well as for
studies of classical and quantum turbulence in superfluid helium.
\end{abstract}

\maketitle
\hyphenation{li-quid dia-phragm quant-ify approx-im-ately}
\section{Introduction}
Much cosmological evidence suggests the existence of dark matter. A
favored candidate for dark matter is the WIMP (Weakly Interacting
Massive Particle), which is the subject of many direct dark matter
searches\protect\cite{GoodmanWitten, Jungman, Bertone, Gaitskell}, in which
nuclear elastic scatters present a potential WIMP signal. Detector technologies using liquefied noble gases have recently become an
important method of direct searches for WIMPs, with several
experiments achieving strong WIMP-nucleon cross-section limits over
the past few years\protect\cite{Angle,Alner,Lebedenko,LArTPCs}. Calibration of such
detectors at low energies (tens of keV) becomes more difficult as
detector sizes are increased due to the self-shielding capability of
large detectors.

\begin{figure}[htb]
\centering
\includegraphics[width=0.35\textwidth, angle=270 ]{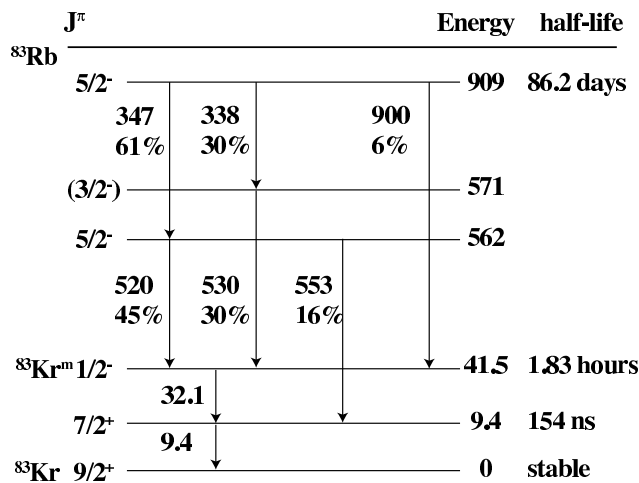}
\caption[Label]{Energy level diagram (in keV) for the $^{83}$Rb decay. Each $^{83}$Rb 
decays
75$\%$ of the time
to the long-lived isomeric $^{83}$Kr$^m$ level 41.5~keV above the ground state, which
subsequently decays in two steps, emitting a 32.1~keV and then a 9.4~keV
conversion electron.} \label{Rb83}
\end{figure}

The doping of a low-energy radioactive source into the active material
of the detector would allow for efficient production of low-energy
events in the fiducial volume of large detectors. The calibration of
the fiducial volume of the XENON10 detector was effected by the
introduction of two activated xenon isotopes produced at Yale
University\protect\cite{ActivatedXe} and shipped to Gran Sasso National Laboratory: $^{129{\rm
m}}$Xe and $^{131{\rm m}}$Xe which emit 236 keV and
164 keV gammas with half-lives of 8.9 and 11.8 days, respectively. 
While these isotopes can be used to fill the detector with a temporary
calibration source, the energies are higher than those expected from
WIMP signals, and their long half-lives prevent their use for frequent
calibration. Another such source is derived from ${^{83}}$Rb, which
decays to $^{83}$Kr$^m$ with a half-life of 86.2 days. The
$^{83}$Kr$^m$ subsequently decays via emission of 32.1 keV and
9.4 keV conversion electrons with a half-life of 1.83
hours~\protect\cite{Kr83mInZeolite} (Figure~\protect\ref{Rb83}). Such a source is not expected to introduce any long-lived
radioisotopes. $^{83}$Kr$^m$ is an important diagnostic tool for
studying the beta spectrum of tritium to determine the neutrino mass, most
recently proposed for use with KATRIN\protect\cite{KATRIN}.

The LUX (Large Underground Xenon) experiment, currently in
fabrication, will be deployed to the Homestake Mine in South Dakota
as a part of the Sanford Underground Science and Engineering
Laboratory (SUSEL). LUX is expected to achieve a WIMP sensitivity at
least two orders of magnitude better than the recent XENON10 and
CDMS-II experiments \protect\cite{Angle,Akerib}. $^{83}$Kr$^m$ can be
used to calibrate the energy scale in the fiducial region, create maps
of detector scintillation and ionization response, and monitor
detector stability.

\section{Experimental Apparatus}

\begin{figure}[htbp]
\centering
\includegraphics[width=0.4\textwidth ]{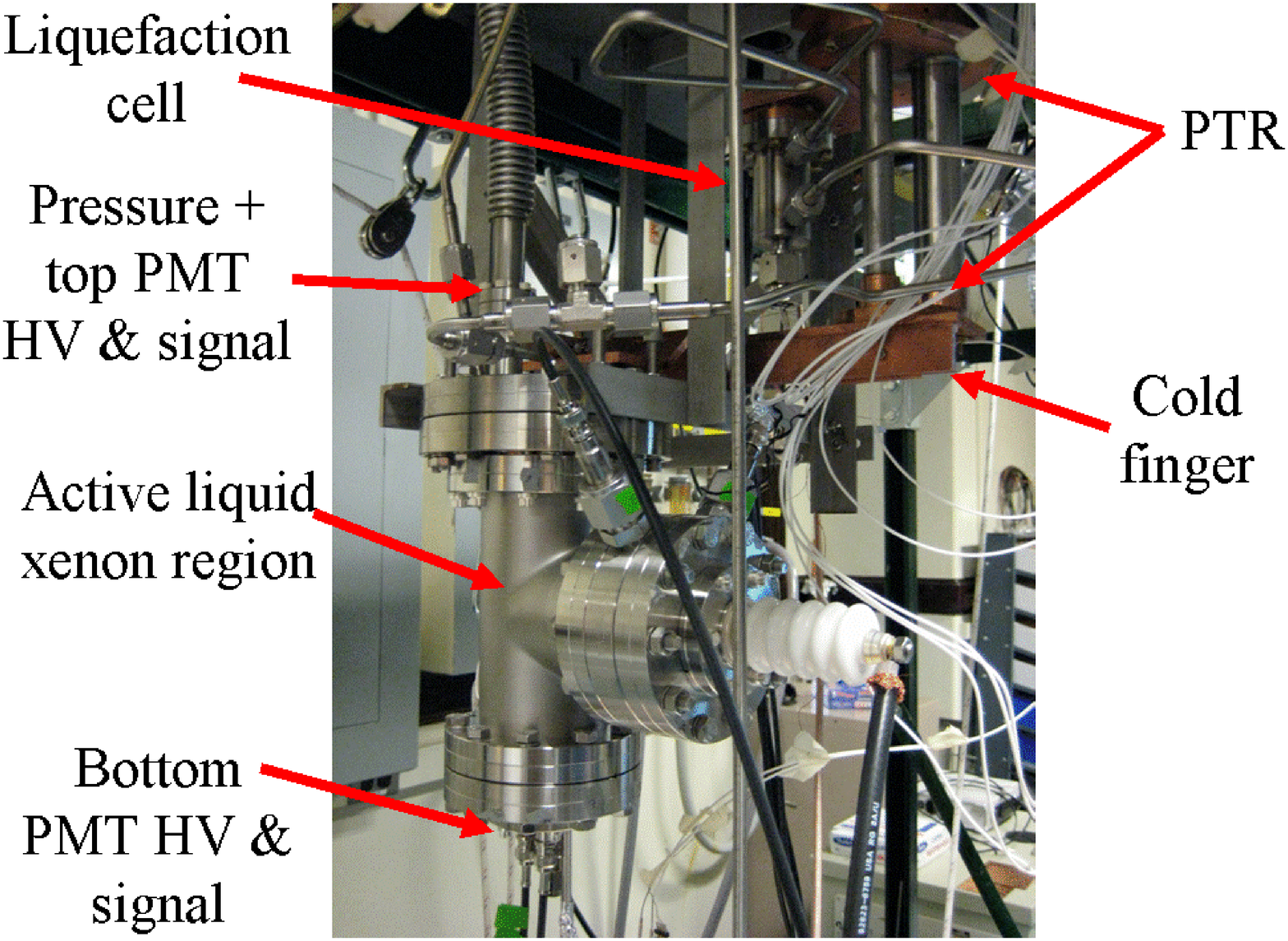}
\caption{(Color online) Photograph of apparatus for $^{83}$Kr$^m$ measurement in liquid xenon.}
\label{pic}
\includegraphics[width=0.4\textwidth ]{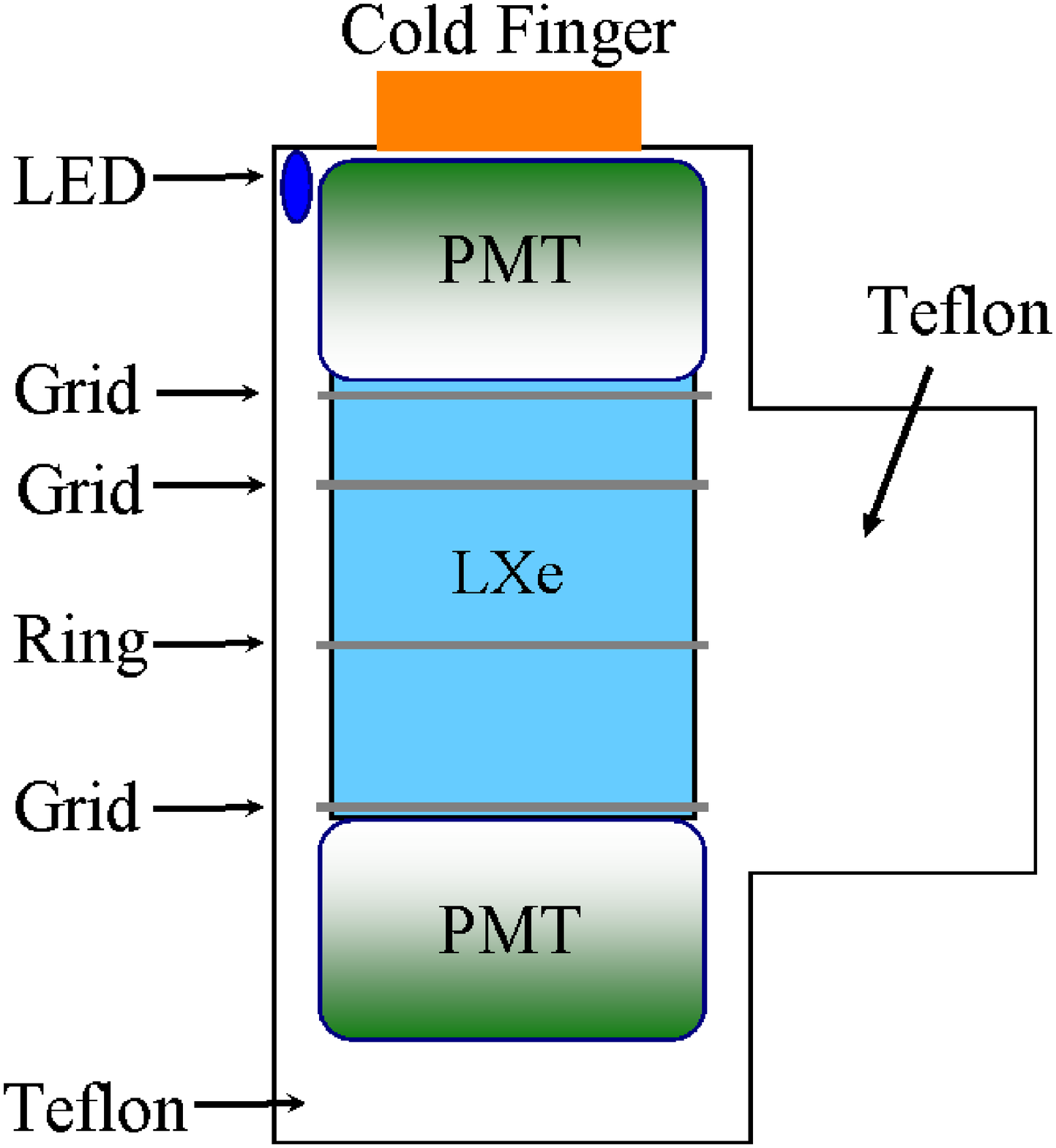}
\caption{(Color online) Schematic of apparatus for $^{83}$Kr$^m$ measurement in liquid xenon.}
\label{setup}
\end{figure}

The liquid xenon detector used to test $^{83}$Kr$^m$ calibration is
located at Yale
University (Fig \ref{pic}). The active liquid xenon target is about
5~cm in
diameter and 2~cm in height, containing 150~g of liquid xenon
surrounded by PTFE (Teflon) for UV scintillation
light reflection and viewed by two Hamamatsu R9869
photomultipliers (PMTs). The R9869 PMT has a
bialkali photocathode with aluminum strip pattern and a quartz
window with a quantum efficiency of 36\% for
xenon
scintillation light at 175~nm. Three stainless steel mesh grids, with 90\% optical
transparency, and a copper ring are
installed for two-phase operation but were not energized in this experiment. The liquid xenon, PMTs, PTFE, and grids are contained in a 11~cm diameter stainless steel cylinder, located in an aluminum vacuum cryostat and
cooled by a pulse-tube
refrigerator (Fig \ref{setup}). The cryogenic system is described elsewhere
\protect\cite{ActivatedXe}.

\begin{figure}[htb]
\centering
\includegraphics[width=0.5\textwidth ]{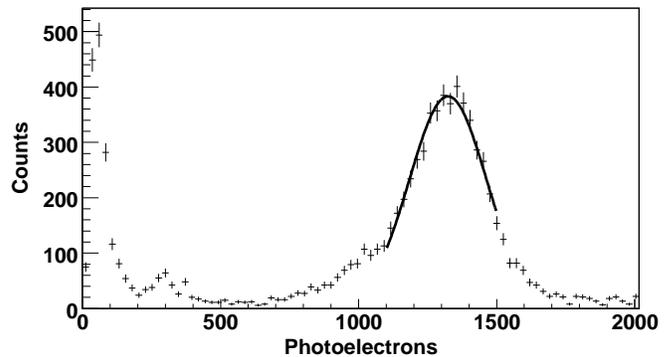}
\caption[Label]{Calibration peak from $^{57}$Co gamma rays, indicating a signal yield of
11.1~pe/keV.}
\label{Co57}
\end{figure}

The gains of the two PMTs are measured from the single photoelectron
(pe) spectra by
using light emitted from an LED inside the liquid xenon detector.
The liquid xenon detector is calibrated with 122~keV and 133~keV gamma rays from a
$^{57}$Co source outside the cryostat. The
scintillation signal yield for
the $^{57}$Co gamma rays in liquid xenon is measured to be 11.1~pe/keV
(Figure~\protect\ref{Co57}). 

\begin{figure}[htbp]
\centering
\includegraphics[width=0.5\textwidth ]{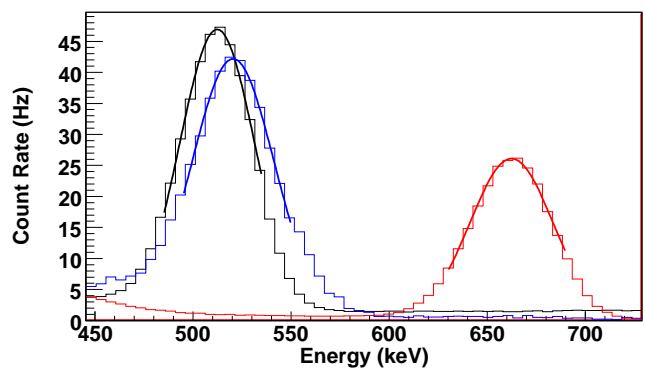}
\caption[Label]{(Color online) After sealing the $^{83}$Kr$^m$ generator, a NaI detector confirms the decay of $^{83}$Rb within the generator. The black curve is a 511~keV gamma ray peak from $^{22}$Na, the red curve is a 662~keV gamma ray peak from $^{137}$Cs, and the blue curve is a peak due to 521~keV, 530~keV and 553~keV gamma rays from $^{83}$Rb decaying in the $^{83}$Kr$^m$ generator.}
\label{Kr83mLoading}
\end{figure}

The $^{83}$Rb source is infused in
2~g of zeolite located in the bottom arm of a VCR cross\protect\cite{Supelco}. The side arms of the cross allow gas to flow
through the chamber,
and filters prevent the introduction of zeolite
into the xenon system. The top arm of the cross allows the zeolite to be
primed with $^{83}$Rb in 1M HCl solution. 700~nCi of $^{83}$Rb was
loaded into a 10~$\mu$l
syringe from a 5~ml glass septum vial, discharged from the
syringe into the
zeolite, sealed in the cross, then baked at 80$^\circ$C for several days. At each step, the amount of  $^{83}$Rb
was measured by
monitoring its combined 521~keV, 530~keV and 553~keV gamma ray emission with a 5 cm
cylindrical NaI
detector (See Figure \protect\ref{Kr83mLoading}).

\begin{figure}[htbp]
\centering
\includegraphics[width=0.22\textwidth, angle=270]{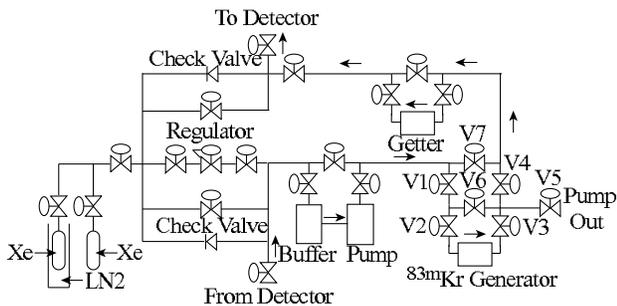}
\caption[Label]{The gas handling system for the xenon detector. The $^{83}$Kr$^m$ 
generator is attached to valves V1 and V4 of the main system and may be separately 
evacuated while the detector is in operation. Closing valve V7 and opening valves V1-V4 
permits the $^{83}$Kr$^m$ to be entrained in xenon flow into the detector.} 
\label{gas_system}
\end{figure}

While the Rb stays attached to the zeolite\protect\cite{Rbtest}, the $^{83}$Kr$^m$
reaches an equilibrium rate of 16~kBq upon introduction to the detector. The $^{83}$Kr$^m$ was doped into the detector by circulating xenon with a diaphragm pump at 2 liters per minute through the $^{83}$Kr$^m$ generator for 5 minutes (Figure \protect\ref{gas_system}). Circulation then bypassed the $^{83}$Kr$^m$ generator, but still passed through the getter. 

\section{Data Analysis and Results}

Events were recorded to computer via a 250~MHz, 12-bit CAEN V1720 waveform digitizer for later analysis. Data acquisition was triggered by coincident PMT pulses, each greater than 1~keV (11~pe). The trigger rate was recorded via a counter and monitored throughout the experiment. After introducing $^{83}$Kr$^m$ into the xenon flow the trigger rate increased for 30 minutes, then began to decay with a half-life consistent with 1.83~h after another 60 minutes. 

The digitized PMT waveforms were analyzed to quantify the response of the detector to $^{83}$Kr$^m$ decays. Each PMT waveform was scanned for a primary and secondary pulse. To find the energy deposited, each pulse was integrated, then divided by the gain of the PMT to determine the number of photoelectrons detected. The number of photoelectrons in each PMT was added and the sum divided by the signal yield determined from the $^{57}$Co calibration. The first pulse was the larger primary pulse corresponding to the 32.1~keV conversion electron. The secondary pulse corresponded to the 9.4~keV conversion electron, emitted following the first conversion electron with a 154~ns half-life. 

\begin{figure}[htb]
\centering
\includegraphics[width=0.5\textwidth ]{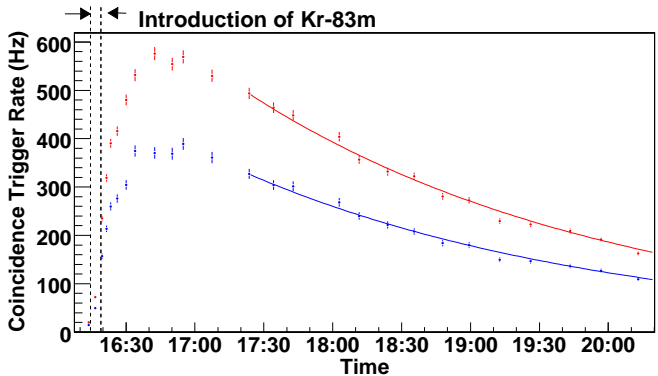}
\caption[Label]{(Color online) The activities of both energy peaks after doping $^{83}$Kr$^m$ into the liquid xenon detector. An exponential was fit to the coincidence trigger rate after it began to decay subsequent to the introduction of the $^{83}$Kr$^m$. The standard ROOT fitting routine was used. The red curve is the 32.1~keV conversion electron with a measured half-life of 1.856~$\pm$~0.036~h, and the blue curve is the 9.4~keV conversion electron
with a measured half-life of 1.846~$\pm$~0.049~h. The decay of both peaks were consistent with the previously measured 1.83~h half-life of $^{83}$Kr$^m$.} 
\label{fig:Kr83mDecay}
\end{figure}

Several cuts were applied to the waveforms. A cut on the asymmetry between the two PMTs was made to optimize the energy resolution. The efficiency of this cut was 50\% for primary pulses and 22\% for secondary pulses. Primary and secondary pulses which occurred close together in time hurt energy resolution. To mitigate this effect, primary pulses with a corresponding secondary pulse less than 4~keV or greater than 14~keV were cut. Similarly, secondary pulses which had a corresponding primary pulse less than 25~keV and greater than 37~keV were cut. The cut to the primary pulse had a 44\% efficiency, while the cut to the secondary pulse had a 56\% efficiency.

\begin{figure}[htb]
\centering
\includegraphics[width=0.45\textwidth 
]{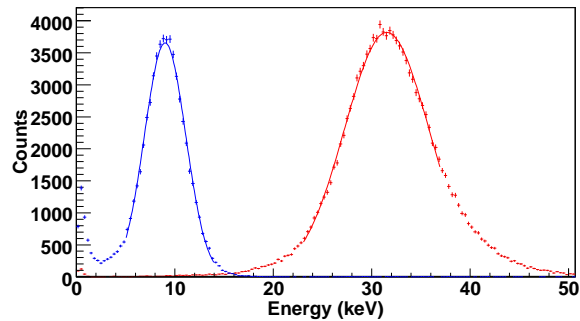}
\caption[Label]{(Color online) Experimental $^{83}$Kr$^m$ peaks measured with liquid xenon scintillation, corresponding to the 9.4~keV 
and 32.1~keV conversion electrons. Noise events create a non-gaussian tail in the 9.4~keV peak below 5~keV, while above 38~keV a non-gaussian tail is seen for the 32.1~keV peak due to 32.1~keV and 9.4~keV conversion electrons which are not be resolved.} \label{fig:Kr83mPeak}
\end{figure}

Combining all of the $^{83}$Kr$^m$ data sets taken over the course of several hours, peaks corresponding to the 9.4~keV and 32.1~keV conversion electrons were fit with a 
Gaussian function using a least-squares fitting method. We found peaks at energies of 31.550~$\pm$~0.019~keV and 9.014~$\pm$~0.010~keV with a linear energy scale derived from the $^{57}$Co calibration 
(Figure~\protect\ref{fig:Kr83mPeak}). Since the scintillation efficiency of liquid xenon may decrease slightly at lower energies \protect\cite{scinteff}, these peaks are consistent with the 32.1~keV and 9.4~keV conversion electrons from $^{83}$Kr$^m$. The 9.4~keV peak had a 23\% resolution ($\sigma/E$), while the 32.1~keV peak had a 14\% resolution.

\begin{figure}[htbp]
\centering
\includegraphics[width=0.5\textwidth 
]{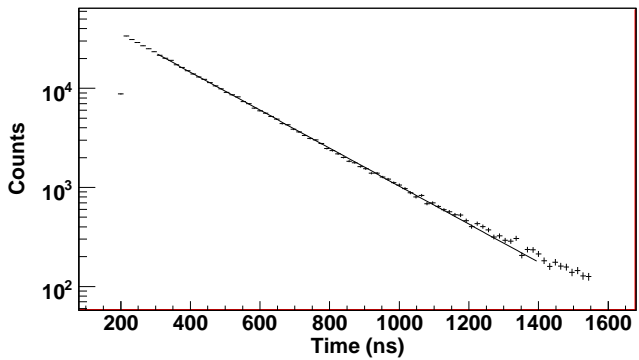}
\caption[Label]{The halflife of $^{83}$Kr$^{\frac{7}{2}^{+}}$ was measured to be 156.94$\pm$0.32~ns, in agreement with previous measurements.} \label{fig:9kevLT}
\end{figure}

We then individually fit twenty six sets of data taken over four hours. The 9.4~keV and 32.1~keV peaks were integrated to find the peak count rate as a function of time. The decay time of each peak was found by fitting an exponential to the count rate data. The half-lives were measured to be 1.856~$\pm$~0.036~h for the 32.1~keV conversion electron and 1.846~$\pm$~0.049~h for the 9.4~keV conversion electron, both consistent with the reported 1.83 half-life of $^{83}$Kr$^m$ (Figure~\protect\ref{fig:Kr83mDecay}). The time between the primary and secondary pulses was fit to an exponential. We found the halflife of $^{83}$Kr$^{\frac{7}{2}^{+}}$ to be 156.94$\pm$0.34~ns, consistent with the previously measured 154.4$\pm$1.1~ns (Figure~\protect\ref{fig:9kevLT}) \protect\cite{9keV_LT}.

\section{Discussion}
Based on the experiment described above, $^{83}$Kr$^m$ may be introduced into low-background liquid xenon detectors 
to characterize the scintillation and ionization response at energies 
similar to those expected from WIMP-nucleon scatters. Typically the newly-purified liquid xenon is 
continuously introduced into the detector, and less pure xenon removed to a room-temperature purification system. With 
such a detector running in a low-background mode, the xenon gas flow 
may be diverted through a section of plumbing containing $\rm 
^{83}Rb$-infused zeolite. The xenon flow will entrain $^{83}$Kr$^m$ atoms, pass through a getter to remove impurities, be liquefied, and enter the detector. With sufficient mixing during their 1.83 hour half-life, 
the $^{83}$Kr$^m$ atoms will spread throughout the active detector 
volume, allowing all parts of the detector to be calibrated for scintillation and ionization response. It is not necessary 
for the $^{83}$Kr$^m$ atoms to spread \textit{evenly} through the 
detector to measure the scintillation and ionization response, though 
an even spread of $^{83}$Kr$^m$ activity would allow a 
calibration of the fiducial mass. 

A pulsed flow of $^{83}$Kr$^m$ atoms would allow the mixing of 
liquid xenon to be visualized in time, as the imaging of individual 
$^{83}$Kr$^m$ decays will indicate the velocity flow of newly 
introduced liquid xenon. The mixing of the liquid xenon flow is 
of interest for liquid xenon detector operation as the degree 
of mixing is important for quantifying the effectiveness of 
purification. With a high degree of mixing, the purity may be 
improved by 1/$e$ at best per volume exchange, but if there is 
little mixing, the degree of purification is limited by the 
efficiency of the purification method. 

Noble liquid detectors using liquid argon, 
liquid neon, and liquid helium might also be calibrated with $^{83}$Kr$^m$, though some fraction of the $^{83}$Kr$^m$ atoms will 
freeze out on detector surfaces. Kr dissolves in liquid argon, while 
in liquid neon and liquid helium, Kr atoms would eventually freeze 
out. In appropriate conditions, the time-scale for Kr freezeout might 
be many times the $^{83}$Kr$^m$ half-life. 

The seeding of $^{83}$Kr$^m$ might also be used in the visualization 
of fluid flow in cryogenic helium. In liquid helium, a decaying $^{83}$Kr$^m$ atom will produce a localized population of approximately 
2000 helium triplet molecules, which may then be imaged using 
laser-induced fluorescence\protect\cite{HeLIF}. At sufficiently low 
temperatures, $^{83}$Kr$^m$ atoms will be trapped by quantized 
vortices in superfluid helium, and so the imaging of 
$^{83}$Kr$^m$ decays through laser-induced fluorescence may permit
the monitoring of vortex dynamics and quantum turbulence 
decay\protect\cite{QuantumTurbHe}. Likewise, introduction of $^{83}$Kr$^m$ 
atoms into a well-defined point in a helium flow at higher 
temperature should allow the visualization of classical turbulence in 
liquid or gaseous helium\protect\cite{ClassicalTurbHe}.

The experiment described above will be augmented 
with additional experiments in which this liquid xenon detector is 
operated in two-phase mode (where ionization is drifted through the 
liquid xenon and detected via proportional scintillation). This 
mode of operation will allow the electron drift length to be 
measured following the introduction of $^{83}$Kr$^m$ into the 
detector to verify that the liquid xenon purity is not 
adversely affected by outgassing of the zeolite. 

\section{Acknowledgments}
The authors would like to recognize the assistance of Jeff 
Ashenfelter and Kevin Charbonneau in the acquisition of the $\rm 
^{83}Rb$ and preparation of the $\rm ^{83}Rb$-infused zeolite. The 
authors would also like to thank Joseph Formaggio, who pointed them 
to the papers of Venos \textit{et al.} describing the preparation of 
$\rm ^{83}Rb$-infused zeolite for the calibration of KATRIN. This 
work was supported by NSF Grant PHY-0800526.


\begin{thebibliography}{99}
\bibitem{Rbtest}This was checked by administering $10^{10}$ more 
$^{\rm nat}$Rb (in aqueous RbCl solution) to a sample of zeolite and 
measuring the Rb escaping into the vacuum with a residual gas 
analyzer while pumping. There was no Rb above baseline, indicating a 
suppression of better than $10^{10}$.
\end{thebibliography}
\end{document}